\documentclass[letterpaper,aps,prd,twocolumn,tightenlines,preprintnumbers,nofootinbib,showkeys,superscriptaddress,longbibliography]{revtex4-1}

\usepackage{lipsum} 

\usepackage{XCharter}
\usepackage[T1]{fontenc}
\usepackage{bm}
\usepackage{bbold}
\usepackage{pifont}
\usepackage{rotating}
\usepackage{fullpage}
\usepackage{amsfonts,shorthand}
\usepackage{amsmath}
\usepackage{slashed}
\usepackage{siunitx}
\usepackage{amssymb}
\usepackage{graphicx}
\usepackage{epic}
\usepackage{eepic}
\usepackage{epsfig}
\usepackage{latexsym}
\usepackage[dvipsnames]{xcolor}
\usepackage[export]{adjustbox}
\usepackage{float}
\usepackage{multirow, makecell}
\usepackage{hyperref}
\usepackage{enumitem}
\hypersetup{colorlinks=true,citecolor=red,linkcolor=NavyBlue,urlcolor=NavyBlue}
\usepackage[caption=false]{subfig}
 
\usepackage{natbib}
\usepackage{relsize}
\usepackage[left=1.65cm,right=1.65cm,top=1.83cm,bottom=1.84cm]{geometry}
\begin{document}
\relscale{1.05}
\title{Interpreting the $\boldsymbol{W}$-Mass and Muon $\boldsymbol{(g_\mu-2)}$ Anomalies\\ within a 2-Higgs Doublet Model }

\author{R. Benbrik}
\email{r.benbrik@uca.ac.ma}
\affiliation{Polydisciplinary Faculty, Laboratory of Fundamental and Applied Physics, Cadi Ayyad University, Sidi Bouzid, B.P. 4162, Safi, Morocco.}
\author{M. Boukidi}
\email{mohammed.boukidi@ced.uca.ma}
\affiliation{Polydisciplinary Faculty, Laboratory of Fundamental and Applied Physics, Cadi Ayyad University, Sidi Bouzid, B.P. 4162, Safi, Morocco.}
\author{B. Manaut}
\email{b.manaut@usms.ac.ma}
\affiliation{Recherche Laboratory in Physics and Engineering Sciences, Team of Modern and Applied Physics, FPBM, USMS -Morocco.}

\begin{abstract}
In this study, we investigate the anomalous magnetic moment of the muon $(g_\mu-2)$ as reported by Fermilab (FNAL), along with the recent measurement of the $W$-boson mass by the CDF-II collaboration. Both findings show significant deviations from the predictions of the Standard Model (SM), hinting at the possibility of new physics. Our focus is on the Type III two-Higgs-doublet model (2HDM), wherein both Higgs doublets couple with all fermions, leading to the induction of flavour-changing neutral currents (FCNCs) at the tree level. Within this framework, we investigate a lepton-flavour-violating (LFV) scenario, aiming to explain both observed anomalies, while satisfying the up-to-date theoretical and experimental constraints.
\end{abstract}
\maketitle
\section{Introduction}
\noindent 

In recent years, a series of experimental results have increasingly suggested the existence of physics beyond the Standard Model (SM). A key example is the measurement of the muon magnetic moment. This measurement, as reported by Fermilab's 'Muon $g-2$' experiment \cite{Muong-2:2023cdq}, indicates a significant deviation from the SM's theoretical predictions. This discrepancy is quantified as $\Delta a_\mu =(25 \pm 4.8) \times 10^{-10}$, constituting a 5.0$\sigma$ discrepancy between theory and experiment.

Concurrently, a major anomaly has been introduced by the CDF-II collaboration's 2022 measurement of the $W$-boson mass \cite{CDF:2022hxs}. The reported value of $m_W = 80.4335 \pm 0.0094$ GeV diverges by 7 standard deviations from the SM's expected value of $m_W = 80.357 \pm 0.006$ GeV \cite{ParticleDataGroup:2020ssz}. This divergence, along with the muon $(g_\mu-2)$ anomaly, suggests the potential for new physics (NP). Various studies have been conducted to explore these anomalies, focusing on the new $W$ mass measurement, the $(g_\mu-2)$ anomaly, or both  \cite{GrillidiCortona:2022kbq, Lu:2022bgw, Heeck:2022fvl, Du:2022brr, Mondal:2022xdy, Chowdhury:2022dps, Bharadwaj:2021tgp, Bahl:2022xzi, Zeng:2022lkk, Ghorbani:2022vtv, Ahn:2022xax, Kawamura:2022uft, Cherchiglia:2022zfy, Bagnaschi:2022qhb, Sakurai:2022hwh, Zheng:2022irz, Lee:2022nqz, Ghoshal:2022vzo, Song:2022xts, Blennow:2022yfm, Du:2022pbp, Herms:2022nhd, Chakraborti:2022vds, Strumia:2022qkt, Popov:2022ldh, Cacciapaglia:2022xih, Atkinson:2021eox, Agrawal:2022wjm, Heo:2022dey, Malaescu:2022qng,  Athron:2022qpo, Cheung:2022zsb, Borah:2023hqw, Fan:2022dck, Borah:2022zim, Zhu:2022tpr, Du:2022fqv, Zhu:2022scj, Arias-Aragon:2022ats, Heckman:2022the, Crivellin:2022fdf, Baek:2022agi, Grimus:2008nb, Alguero:2022est, Carpenter:2022oyg, Borah:2022obi, Abdallah:2022shy, Arcadi:2022dmt, Han:2022juu, Atkinson:2022pcn, Liu:2022jdq, Cheng:2022jyi, Arora:2022uof, DiLuzio:2022xns, Almeida:2022lcs, Yuan:2022cpw, Babu:2022pdn, FileviezPerez:2022lxp, Yang:2022gvz, Ahn:2022xeq, Addazi:2022fbj, Kanemura:2022ahw, Cao:2022mif, Cheng:2022aau, Krasnikov:2022xsi, Peli:2022ybi, Biekotter:2022abc, Athron:2022isz, Barman:2022iwj, Chowdhury:2022moc, Bhaskar:2022vgk, Nagao:2022oin, Wilson:2022gma, Zhang:2022nnh, Song:2022jns, Bahl:2022gqg, Borah:2023hqw, Li:2023mrw, Liu:2023kjc, deGiorgi:2022xhr, Chen:2023mep, Crivellin:2023xbu, Mishra:2023cjc, Chen:2023eof, Arcadi:2023qgf, Cacciapaglia:2022evm, Arora:2022uof, Chakrabarty:2022voz, Kim:2022zhj, Chowdhury:2022dps, He:2022zjz, Dcruz:2022dao, Kim:2022xuo, Kim:2022hvh, Botella:2022rte, Zhou:2022cql, Tang:2022pxh, Chakraborti:2022wii, Ali:2021kxa, Chakraborti:2021kkr, deJesus:2023som}.

In response to these anomalies, the two-Higgs-doublet model (2HDM) has gained attention as a promising framework for new physics beyond the SM. The 2HDM introduces an additional SU(2)$_L$ Higgs doublet, leading to the prediction of two neutral CP-even Higgs bosons ($h$ and $H$), a neutral pseudoscalar ($A$), and charged Higgs bosons ($H^\pm$). This model is particularly notable for its potential to correct gauge boson masses and address the muon $(g_\mu-2)$ anomaly through specific lepton interactions.

In this work, we explore a lepton-flavour-violating $\mu-\tau$ (LFV) scenario \cite{Crivellin:2013wna} within a generic 2HDM (Type-III) \cite{Hernandez-Sanchez:2012vxa,Benbrik:2022azi,Benbrik:2021wyl}, incorporating Flavour Changing Neutral Currents (FCNCs) in the Yukawa sector. Our aim is to simultaneously explain both the $m_W$ and $(g_\mu-2)$ anomalies while adhering to current theoretical and experimental constraints.

The paper is organized as follows: Section \ref{sec1} introduces the 2HDM-III model. Section \ref{sec2} analyses the $W$-boson mass and the muon $(g_\mu-2)$ anomalies, respectively, in light of relevant theoretical and experimental constraints. Section \ref{sec3} concludes with a discussion of our findings.

\section{General 2HDM}\label{sec1}
The 2HDM serves as one of the most straightforward extensions of the SM. It comprises two complex doublets of Higgs fields, denoted as $\Phi_i$ ($i = 1, 2$), each with a hypercharge of $Y = +1$. The scalar potential, invariant under the SU(2)$_L \otimes$U(1)$_Y$ gauge symmetry, can be expressed as\cite{Branco:2011iw}:

\begin{align}
\mathcal{V} &= m_{11}^2 \Phi_1^\dagger \Phi_1+ m_{22}^2\Phi_2^\dagger\Phi_2 - \left[m_{12}^2
\Phi_1^\dagger \Phi_2 + \rm{H.c.}\right]  ~\nonumber\\&+ \lambda_1(\Phi_1^\dagger\Phi_1)^2 +
\lambda_2(\Phi_2^\dagger\Phi_2)^2 +
\lambda_3(\Phi_1^\dagger\Phi_1)(\Phi_2^\dagger\Phi_2)  ~\nonumber\\ &+
\lambda_4(\Phi_1^\dagger\Phi_2)(\Phi_2^\dagger\Phi_1) +
\frac12\left[\lambda_5(\Phi_1^\dagger\Phi_2)^2 +\rm{H.c.}\right] 
 ~\nonumber\\&+\left\{\left[\lambda_6(\Phi_1^\dagger\Phi_1)+\lambda_7(\Phi_2^\dagger\Phi_2)\right]
(\Phi_1^\dagger\Phi_2)+\rm{H.c.}\right\} \label{C2HDMpot}
\end{align}

The hermiticity of this potential implies that the parameters $m_{11}^2$, $m_{22}^2$, and $\lambda_{1,2,3,4}$ are real. In contrast, $\lambda_{5,6,7}$ and $m_{12}^2$ can be complex, although they are considered real in the CP-conserving version of the 2HDM. Notably, the $\lambda_{6,7}$ terms, while influencing the $h\gamma\gamma$ process through triple Higgs couplings, have a minimal effect in this study and are thus set to zero. This simplification leaves the model with seven independent parameters, reduced to six in our analysis with the assumption of $H$ as the observed SM-like Higgs boson with a mass of 125 GeV.

The Yukawa sector of the 2HDM involves general scalar-to-fermion couplings, expressed as:

\begin{align}
-{\cal L}_Y &= \bar Q_L Y^u_1 U_R \tilde \Phi_1 + \bar Q_L Y^{u}_2 U_R
\tilde \Phi_2  + \bar Q_L Y^d_1 D_R \Phi_1 
\nonumber \\&+ \bar Q_L Y^{d}_2 D_R \Phi_2 
+  \bar L Y^\ell_1 \ell_R \Phi_1 + \bar L Y^{\ell}_2 \ell_R \Phi_2 + H.c. 
\label{eq:Yu}
\end{align}
Before Electro-Weak Symmetry Breaking (EWSB), the Yukawa matrices $Y^{f}{1,2}$, which govern the interactions between the Higgs fields and fermions, are arbitrary $3\times 3$ matrices. In this state, fermions do not yet represent physical eigenstates. This allows us the flexibility to choose diagonal forms for the matrices $Y^u_1$, $Y^d_2$, and $Y^\ell_2$. Specifically, we can set $Y^u_1 = \mathrm{diag}(y^u_1, y^u_2, y^u_3)$ and $Y^{d,\ell}_{2} = \mathrm{diag}(y^{d,\ell}_1, y^{d,\ell}_2, y^{d,\ell}_{3})$.

In our study, we focus on the Type-III 2HDM. This variant does not impose a global symmetry on the Yukawa sector nor enforces alignment in flavour space. Instead, we adopt the Cheng-Sher ansatz \cite{Cheng:1987rs, Diaz-Cruz:2004wsi}, which posits a specific flavour symmetry in the Yukawa matrices. Under this assumption, FCNC effects are proportional to the masses of the fermions and dimensionless real parameters, with the relationship $\tilde{Y}{ij} \propto \sqrt{m_i m_j}/v \cdot \chi{ij}$.	
After EWSB, the Yukawa Lagrangian is expressed in terms of the mass eigenstates of the Higgs bosons. It can be represented as follows:
\begin{widetext}
\begin{align}
-{\cal L}^{III}_Y  &= \sum_{f=u,d,\ell} \frac{m^f_j }{v} \times\left( (\xi^f_h)_{ij}  \bar f_{Li}  f_{Rj}  h + (\xi^f_H)_{ij} \bar f_{Li}  f_{Rj} H - i (\xi^f_A)_{ij} \bar f_{Li}  f_{Rj} A \right)\nonumber\\  &+ \frac{\sqrt{2}}{v} \sum_{k=1}^3 \bar u_{i} \left[ \left( m^u_i  (\xi^{u*}_A)_{ki}  V_{kj} P_L+ V_{ik}  (\xi^d_A)_{kj}  m^d_j P_R \right) \right] d_{j}  H^+ \nonumber\\  &+ \frac{\sqrt{2}}{v}  \bar \nu_i  (\xi^\ell_A)_{ij} m^\ell_j P_R \ell_j H^+ + H.c.\, \label{eq:Yukawa_CH}
\end{align} 	
\end{widetext}
Here, $V_{kj}$ represents the Cabibbo-Kobayashi-Maskawa (CKM) matrix, while the specific reduced Yukawa couplings are elaborated in Table \ref{coupIII}, with expressions defined in relation to the mixing angle $\alpha$, $\tan\beta$, and the independent parameters $\chi_{ij}^f$.

\begin{table*}[t]
	\begin{center}
		\setlength{\tabcolsep}{13pt}
		\renewcommand{\arraystretch}{1} %
		\begin{tabular}{c|c|c|c} \hline\hline 
			$\phi$  & $(\xi^u_{\phi})_{ij}$ &  $(\xi^d_{\phi})_{ij}$ &  $(\xi^\ell_{\phi})_{ij}$  \\   \hline
			$h$~ 
			& ~ $  \frac{c_\alpha}{s_\beta} \delta_{ij} -  \frac{c_{\beta-\alpha}}{\sqrt{2}s_\beta}  \sqrt{\frac{m^u_i}{m^u_j}} \chi^u_{ij}$~
			& ~ $ -\frac{s_\alpha}{c_\beta} \delta_{ij} +  \frac{c_{\beta-\alpha}}{\sqrt{2}c_\beta} \sqrt{\frac{m^d_i}{m^d_j}}\chi^d_{ij}$~
			& ~ $ -\frac{s_\alpha}{c_\beta} \delta_{ij} + \frac{c_{\beta-\alpha}}{\sqrt{2}c_\beta} \sqrt{\frac{m^\ell_i}{m^\ell_j}}  \chi^\ell_{ij}$ ~ \\
			$H$~
			& $ \frac{s_\alpha}{s_\beta} \delta_{ij} + \frac{s_{\beta-\alpha}}{\sqrt{2}s_\beta} \sqrt{\frac{m^u_i}{m^u_j}} \chi^u_{ij} $
			& $ \frac{c_\alpha}{c_\beta} \delta_{ij} - \frac{s_{\beta-\alpha}}{\sqrt{2}c_\beta} \sqrt{\frac{m^d_i}{m^d_j}}\chi^d_{ij} $ 
			& $ \frac{c_\alpha}{c_\beta} \delta_{ij} -  \frac{s_{\beta-\alpha}}{\sqrt{2}c_\beta} \sqrt{\frac{m^\ell_i}{m^\ell_j}}  \chi^\ell_{ij}$ \\
			$A$~  
			& $ \frac{1}{t_\beta} \delta_{ij}- \frac{1}{\sqrt{2}s_\beta} \sqrt{\frac{m^u_i}{m^u_j}} \chi^u_{ij} $  
			& $ t_\beta \delta_{ij} - \frac{1}{\sqrt{2}c_\beta} \sqrt{\frac{m^d_i}{m^d_j}}\chi^d_{ij}$  
			& $t_\beta \delta_{ij} -  \frac{1}{\sqrt{2}c_\beta} \sqrt{\frac{m^\ell_i}{m^\ell_j}}  \chi^\ell_{ij}$ \\ \hline \hline 
		\end{tabular}
	\end{center}
	\caption {Yukawa couplings of the neutral Higgs bosons $h$, $H$, and $A$ to the quarks and leptons in the 2HDM Type-III.} 
	\label{coupIII}
\end{table*}
\section{${m_W}$ and $(g_\mu-2)$ in the 2HDM}\label{sec2}

Within the framework of the 2HDM, the $W$ boson mass stands out as a pivotal Electroweak Precision Observable (EWPO). Its significance is underscored by the precision achieved in experimental measurements. This mass can be expressed as a function of the oblique parameters $S$, $T$, and $U$. These parameters quantify deviations from SM predictions due to new physics influences, as elaborated in \cite{Peskin:1990zt,Peskin:1991sw}.

Specifically in the 2HDM, the contributions to the oblique parameters $S$, $T$, and $U$ predominantly stem from additional Higgs boson loops and altered interactions of the SM-like Higgs bosons. This results in a modified expression for the $W$-boson mass, as given by \cite{Grimus:2008nb}:
\begin{widetext}
\begin{equation}
m_W^2 = m_W^2|_{SM}\Bigg[1 + \frac{\alpha(M_Z)}{c_W^2 - s_W^2}\left(-\frac{1}{2}S + c_W^2T + \frac{c_W^2 - s_W^2}{4s_W^2}U\right)\Bigg],
\end{equation}
\end{widetext}
where $c_W = \cos \theta_W$ and $s_W = \sin \theta_W$ are the cosine and sine of the Weinberg angle, respectively. In many scenarios involving new physics, the $U$ parameter's contribution is typically minor compared to $S$ and $T$. Consequently, we assume $U = 0$ for simplicity in our analysis.

Furthermore, changes in the $W$-boson mass impact the effective weak mixing angle $\sin^2\theta_{\mathrm{eff}}$, crucial in electroweak theory. This angle, when adjusted for new physics contributions, can be calculated as~\cite{Biekotter:2022abc}:

\begin{equation}
\sin^2\theta_{\mathrm{eff}} = \sin^2\theta_{\mathrm{eff}}|_{SM} - \alpha\frac{c_W^2s_W^2}{c_W^2 - s_W^2}(T - \frac{s_W^2}{c_W^2}S + \frac{1}{2}U),
\end{equation}

with established SM values being $m_W|_{SM} = 80.357$ GeV and $\sin^2\theta^{\mathrm{eff}}|_{SM} = 0.231532$.

In addressing the muon anomalous magnetic moment $(g_\mu-2)$, the Type-III 2HDM notably contributes through one-loop diagrams featuring the Higgs bosons $h$, $H$, and $A$, as well as lepton flavour-violating (LFV) interactions. The contribution to the anomalous magnetic moment, $\Delta a_\mu$, can be approximated as \cite{Benbrik:2015evd,Assamagan:2002kf,Davidson:2010xv}:

\begin{equation}
\Delta a_\mu \approx \frac{m_\mu m_\tau X_{23}^\ell X_{32}^\ell}{8\pi^2 c_\beta^2}Z_{\phi},
\end{equation}

where $X_{23}^\ell = \sqrt{m_\mu m_\tau}/v\chi_{23}^\ell$ and $Z_{\phi}$ is defined as:

\begin{align}
Z_\phi &= \frac{c_{\beta\alpha}^2\left(\ln\left( m_h^2/m_\tau^2\right)-\frac{3}{2}\right)}{m_h^2} + \frac{s_{\beta\alpha}^2\left(\ln\left( m_H^2/m_\tau^2\right)-\frac{3}{2}\right)}{m_H^2} \nonumber \\&- \frac{\ln\left(m_A^2/m_\tau^2\right)-\frac{3}{2}}{m_A^2}.
\end{align}

In this analysis, the terms $X_{23}^\ell$ and $X_{32}^\ell$ are treated symmetrically, reflecting the symmetric nature of the $\mu-\tau$ interactions within this model {More details can be found in\cite{Benbrik:2015evd}}. 

\subsection{Theoretical and experimental constraints}
In our work, we employ a diverse set of theoretical and experimental constraints that must be met to establish a viable model.
\begin{itemize}
	\item \textbf{Unitarity}: Scattering processes in the 2HDM must satisfy the requirement of unitarity \cite{Kanemura:1993hm}.
	\item \textbf{Perturbativity}: The quartic couplings of the scalar potential should be perturbative, with absolute values satisfying $|\lambda_i| < 8\pi$ for each $i=1,...,5$ \cite{Branco:2011iw}.
	\item \textbf{Vacuum Stability}: The scalar potential must be bounded from below and positive in all field directions. Therefore, the parameter space must satisfy the conditions \cite{Barroso:2013awa,Deshpande:1977rw}:
	\begin{align}
	&\lambda_1 > 0, \quad \lambda_2 > 0, \quad \lambda_3 > -\sqrt{\lambda_1\lambda_2}, \nonumber\\
	&\lambda_3+\lambda_4-|\lambda_5| > -\sqrt{\lambda_1\lambda_2}.
	\end{align}
	\item \textbf{SM-like Higgs Boson Discovery}: The compatibility of the SM-like scalar with the observed Higgs boson is tested. The relevant quantities calculated with \texttt{HiggsSignals-3} \cite{Bechtle:2020pkv,Bechtle:2020uwn} via \texttt{HiggsTools} \cite{Bahl:2022igd}  must satisfy the measurements at 95\% confidence level (C.L.).
	\item \textbf{BSM Higgs Boson Exclusions}: Exclusion limits at 95\% C.L. from direct searches for Higgs bosons at LEP, Tevatron, and LHC are taken into account using \texttt{HiggsBounds-6} \cite{Bechtle:2008jh,Bechtle:2011sb,Bechtle:2013wla,Bechtle:2015pma} via \texttt{HiggsTools}, and includes considerations of lepton violation processes, such as the decay $h\to\tau\nu$ \cite{ATLAS:2019pmk}.
	\item \textbf{LEP Measurement Compatibility}: For a given parameter point in our model to be allowed, the corresponding partial width $\Gamma(Z \to hA)$ must be within 2$\sigma$ experimental uncertainty of the measurement when it is kinematically open. The total width of the $Z$ boson is taken to be $\Gamma_Z = 2.4952 \pm 0.0023$ GeV \cite{ALEPH:2005ab}.
	
	\item
	{\bf $B$-physics observables}:The constraints from $B$-physics observables are implemented using the code \texttt{SuperIso\_v4.1} \cite{Mahmoudi:2008tp} as described in Ref. \cite{Benbrik:2022azi}. The relevant experimental measurements used are as follows:
	
{	\begin{enumerate}
		\item ${\cal BR}(\overline{B}\to X_s\gamma)|_{E_\gamma<1.6\mathrm{~GeV}}$ $\left(3.32\pm0.15\right)\times 10^{-4}$ \cite{HFLAV:2016hnz}
		\item${\cal BR}(B^+\to \tau^+\nu_\tau)$ $\left(1.06\pm0.19\right)\times 10^{-4}$ \cite{HFLAV:2016hnz}
		\item${\cal BR}(D_s\to \tau\nu_\tau)$ $\left(5.51\pm0.18\right)\times 10^{-2}$ \cite{HFLAV:2016hnz}
		\item${\cal BR}(B_s\to \mu^+\mu^-)$ (LHCb) $\left(3.09^{+0.46}_{-0.43}\right)\times 10^{-9}$ \cite{LHCb:2021awg,LHCb:2021vsc}
		\item	${\cal BR}(B_s\to \mu^+\mu^-)$ (CMS) $\left(3.83^{+0.38}_{-0.36}\right)\times 10^{-9}$ \cite{CMS:2022mgd}
		\item	${\cal BR}(B^0\to \mu^+\mu^-)$ (LHCb) $\left(1.2^{+0.8}_{-0.7}\right)\times 10^{-10}$ \cite{LHCb:2021awg,LHCb:2021vsc}
		\item	${\cal BR}(B^0\to \mu^+\mu^-)$ (CMS) $\left(0.37^{+0.75}_{-0.67}\right)\times 10^{-10}$ \cite{CMS:2022mgd}
	\end{enumerate} }
\end{itemize}

\section{Numerical Results}\label{sec3}

Considering both the Normal Scenario (NS) and Inverted Scenario (IS), where the observed 125 GeV Higgs boson at the LHC is assigned to either the $\mathcal{CP}$-even state $h$ (NS) or $H$ (IS), we perform a random scan of the parameter space of the 2HDM Type-III using the public code \texttt{2HDMC-1.8.0}. The ranges for the scanned parameters are shown in Table \ref{tab2}. Each scanned point must satisfy the theoretical and experimental constraints discussed earlier.
	\begin{table*}[htp!]
	\centering
		\setlength{\tabcolsep}{0.1pt}
\renewcommand{\arraystretch}{1.10} %
		\begin{tabular}{c|c|c|c|c|c|c|c|c|c}\hline\hline
			&	$m_h~[\mathrm{GeV}]$&$m_H~[\mathrm{GeV}]$&$m_A~[\mathrm{GeV}]$&	$m_{H^\pm}~[\mathrm{GeV}]$& $c_{\beta-\alpha}$&$\tan\beta$&$Z_5$&$Z_4$&$\chi_{ij}^{f,\ell}$ \\\hline
			NS	&	$125.09$&$[126;\,1000]$&$[80;\,1000]$&$[80;\,1000]$& $[0;\,0.4]$&$[2;\,100]$&$(1/v)(m_H^2s_{\beta-\alpha}^2+m_hc_{\beta-\alpha}^2-m_A^2)$&$2/v(m_A^2-m_{H^\pm}^2)+Z_5$&$[-10;\,10]$\\\hline
			IS	&	$[20;\,120]$&$125.09$&$[80;\,1000]$&$[80;\,1000]$& $[0.6;\,1]$&$[2;\,100]$&$(1/v)(m_H^2s_{\beta-\alpha}^2+m_hc_{\beta-\alpha}^2-m_A^2)$&$2/v(m_A^2-m_{H^\pm}^2)+Z_5$&$[-10;\,10]$\\\hline\hline
		\end{tabular}

	\caption{2HDM input parameters.} \label{tab2}
\end{table*}

\begin{figure*}[t!]	
	\centering
	\mbox{\includegraphics[height=20cm,width=18cm]{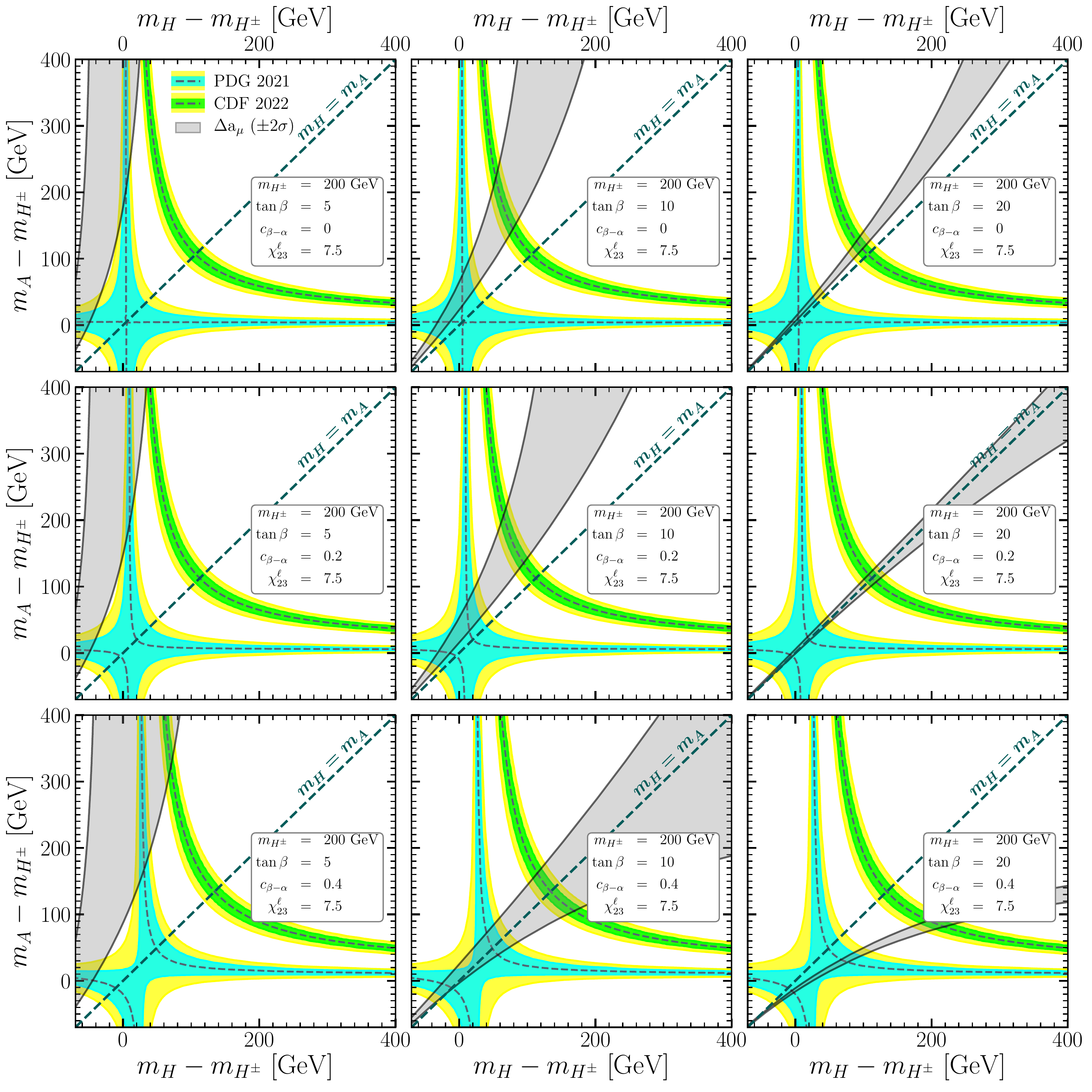}}
	\caption{The allowed regions ($1$-and $2\sigma$)  by the oblique parameter fit for 2HDM Type-III in the plane of $m_H-m_{H^\pm}$ versus $m_A-m_{H{^\pm}}$. The gray area shows the $2\sigma$ allowed region by $\Delta a_\mu$. }\label{fig1}
\end{figure*}
	\subsection{Part I: Inverted scenario}

In this scenario, we identify the light $\mathcal{CP}$-even Higgs boson, $h$, as the signal observed at the LHC, fixing its mass at $m_h = 125.09$ GeV. We initially conduct a targeted scan focused on mass splitting, with specific values set for $\tan\beta$, $c_{\beta-\alpha}$, and $\chi^{f,\ell}_{ij}$.
\begin{figure}[htp!]
	\centering
	\includegraphics[height=8.5cm,width=8.0cm]{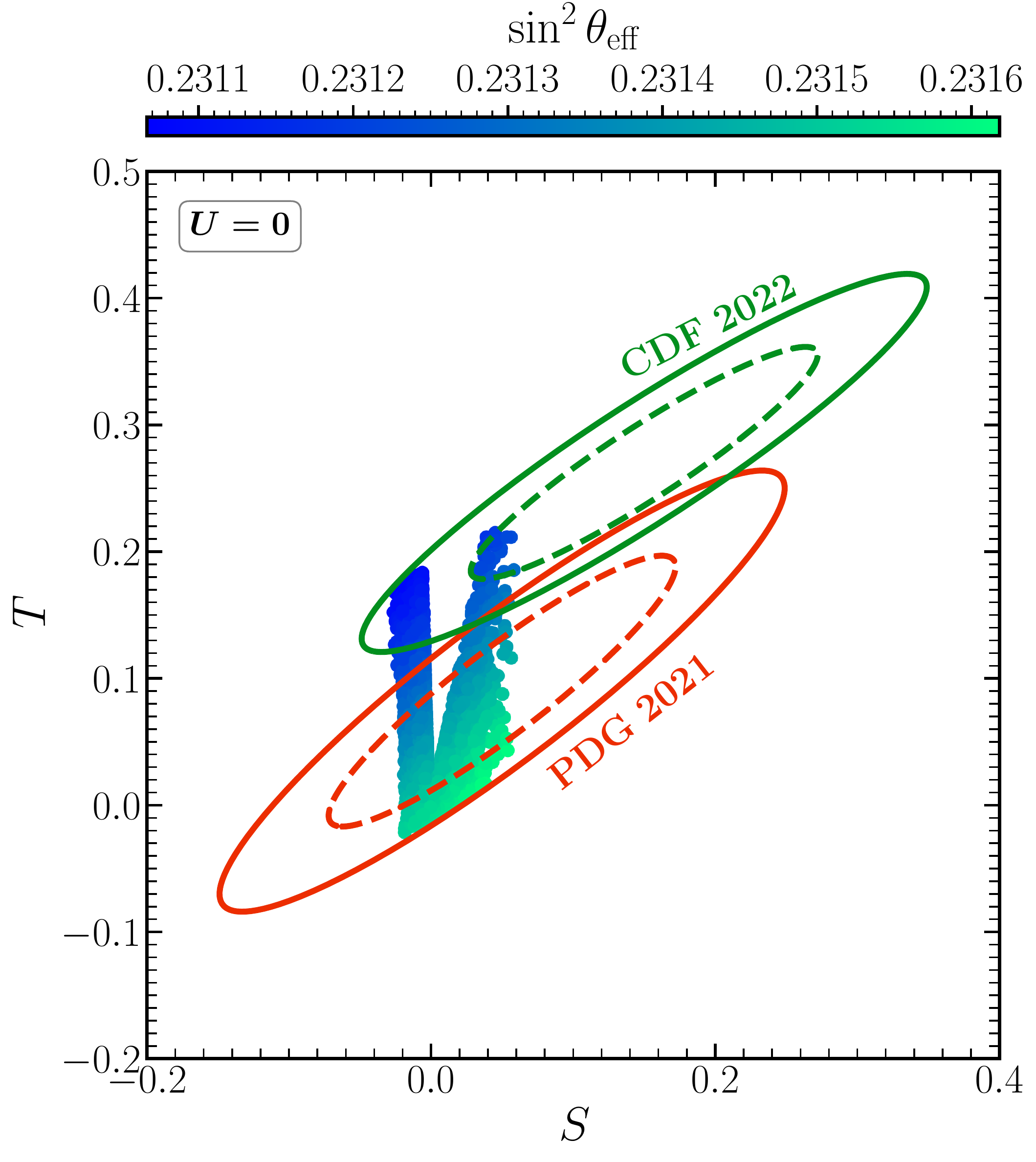}
	\caption{Allowed points by all constraints including $g-2$ superimposed onto the fit limits in the
		$( S, ~ T)$ plane from EWPO data at 68\% (solid lines) and 95\% dashed lines Confidence Level (CL) with a correlation of 92(93)\%  PDG(CDF) with $U=0$. The colour coding indicates the value $\sin^2\theta_{\mathrm{eff}}$.}\label{fig2}
\end{figure}
\begin{figure}[htp!]
	\centering
	\includegraphics[height=8.5cm,width=8.0cm]{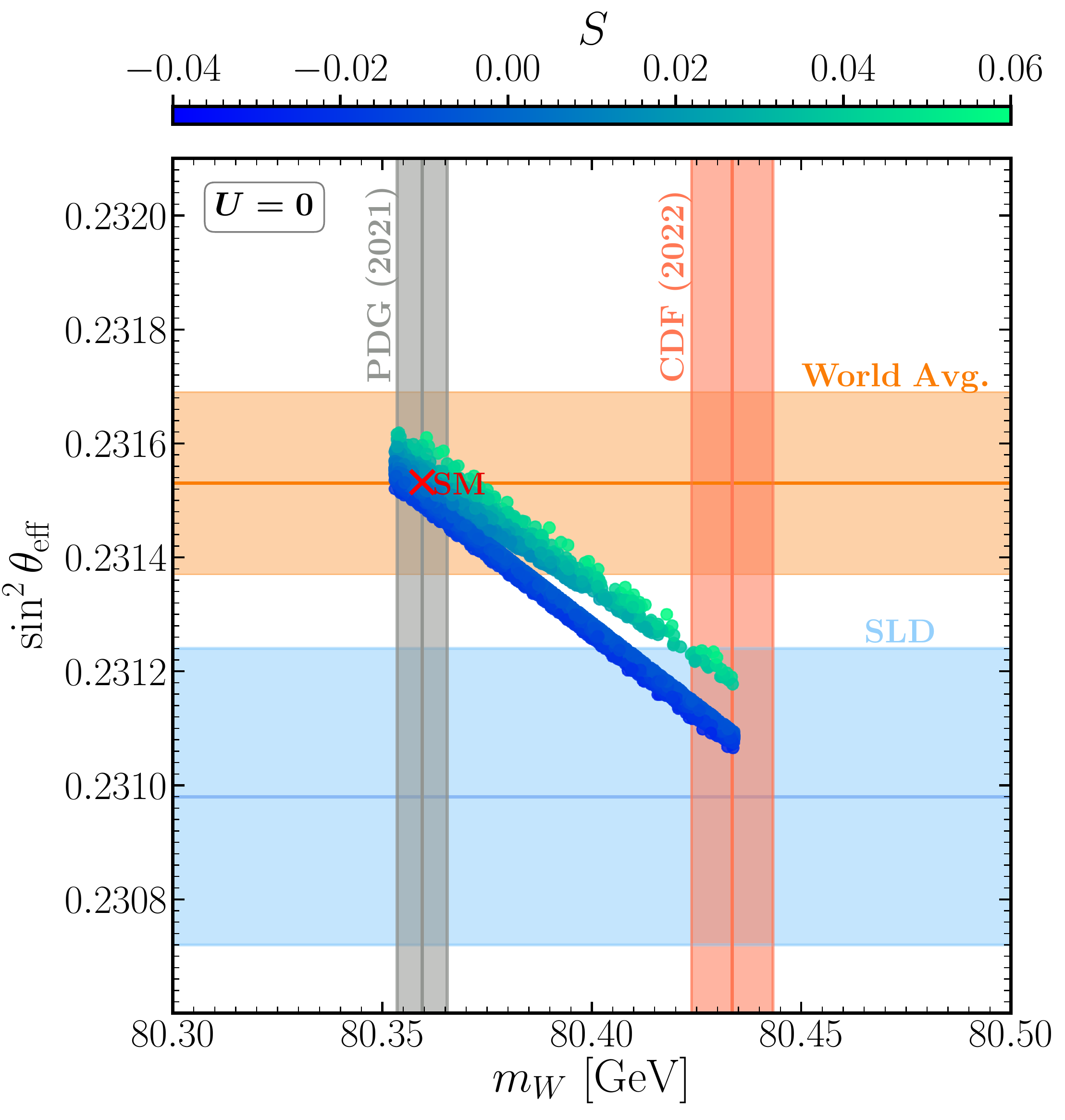}
	
	\caption{Same points as in Fig. \ref{fig2} mapped into the $(m_W , \sin^2 \theta_{\mathrm{eff}})$ plane. The colour coding indicates the value of $S$. The light red line represent the CDF-II measured $W$ boson mass and its 1$\sigma$ range (light red band). The gray line represent the PDG value for the $W$ boson mass with the gray band showing the 1$\sigma$ range. The orange line  represent the world averaged value $0.23153 \pm 0.00016$ \cite{ParticleDataGroup:2020ssz,ALEPH:2005ab} with its 1$\sigma$ range (orange band), while the light blue line illustrates the SLD measured value $0.23098 \pm 0.00026$ \cite{ALEPH:2005ab} of $\sin^2 \theta_{\mathrm{eff}}$ with the light blue band showing the 1$\sigma$ range. The red cross indicates the SM prediction.}\label{fig3}
\end{figure}
\begin{figure}[htp!]
	\centering
	\includegraphics[height=8.5cm,width=8.0cm]{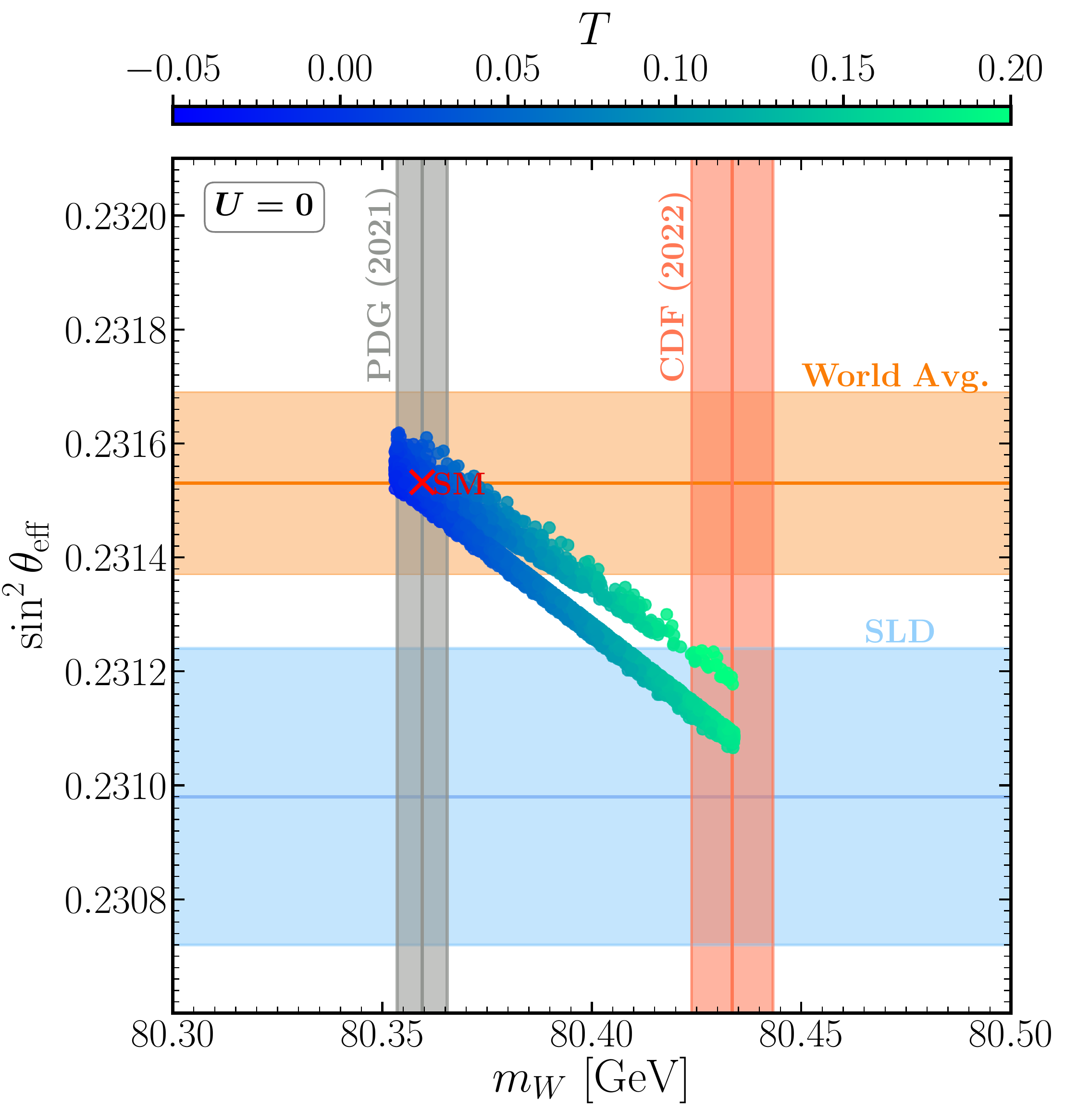}
	\caption{Same as in Fig. \ref{fig3} but with the $T$ paramete in the colour bar.}\label{fig4}
\end{figure}
The results of this targeted scan are depicted in Fig.~\ref{fig1}, which displays the allowed regions within the 1- and 2-$\sigma$ levels as determined by the $S$ and $T$ parameters in the $((m_H - m_{H^\pm}), (m_A - m_{H^\pm}))$ plane. The PDG  data are represented in light blue and yellow for the 1$\sigma$ and 2$\sigma$ intervals, respectively, while the CDF  results appear in green and yellow. Grey contours indicate compatibility with the 2$\sigma$ muon $(g_\mu-2)$ findings. The plots reveal that scenarios with degenerate masses for $m_{H^\pm}$, $m_A$, and $m_H$ are strongly disfavoured by the new CDF data. However, regions exist where mass degeneracy between $m_H$ and $m_A$ aligns with both the muon $(g_\mu-2)$ anomaly and the CDF reported $m_W$ value. This agreement is particularly feasible with parameter sets like $\tan\beta = 20$, $c_{\beta-\alpha} = 0.2$, and $\tan\beta = 10$, $c_{\beta-\alpha} = 0.4$.

Moving to the comprehensive scan detailed in Table~\ref{tab2}, Fig~\ref{fig2} shows the parameter points that meet all discussed theoretical and experimental constraints, including those aligned with the latest muon $(g_\mu-2)$ measurement at the 2$\sigma$ level. The plot's colour coding reflects $\sin^2\theta_{\mathrm{eff}}$ values, with solid and dashed ellipses outlining the 1$\sigma$ and 2$\sigma$ confidence regions for $\chi_{S,T}^2$, in red for PDG and green for CDF constraints.

The plot demonstrates the existence of viable points within the 2$\sigma$ range that account for the muon $(g_\mu-2)$ anomaly while adhering to other relevant theoretical and experimental constraints. These points also fall within the 1$\sigma$ ellipse for the CDF  $W$-boson mass measurement, suggesting that the 2HDM Type-III can aptly fit the CDF(2022) $W$-boson mass data and explain the muon $(g_\mu-2)$ anomaly.
\begin{figure}[htp!]
	\centering
	\includegraphics[height=8.2cm,width=8.15cm]{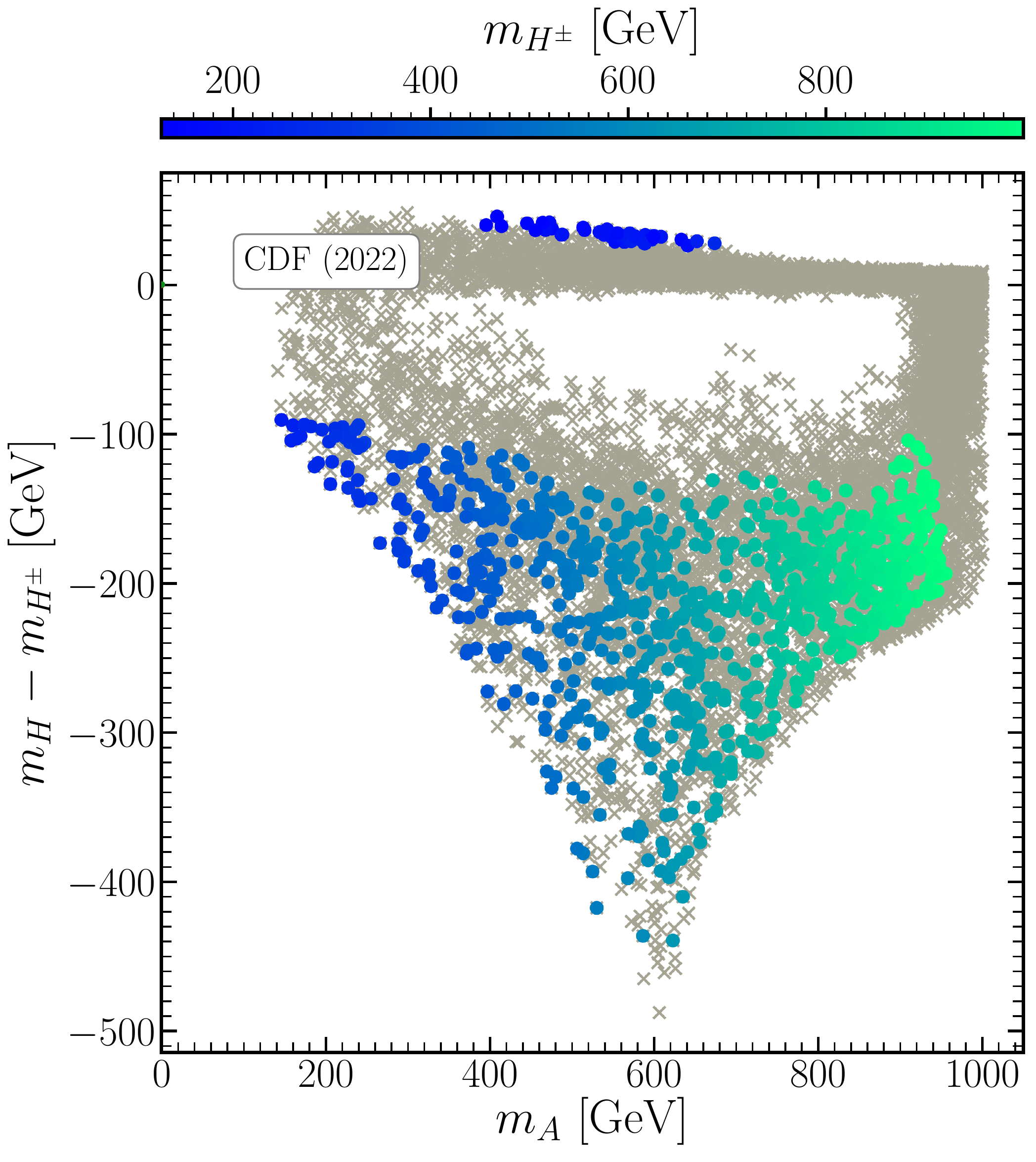}
	\caption{Same points as in Fig. \ref{fig2} mapped into the $(m_A , m_H-m_{H^\pm})$ plane. The colour coding indicates the value of $m_{H^\pm}$ using the CDF 2022 data. The gray crosses  represent the points allowed only by theoretical and experimental constraints.}\label{fig5}
\end{figure}
\begin{figure}[htp!]
	\centering
	\includegraphics[height=8.2cm,width=8.15cm]{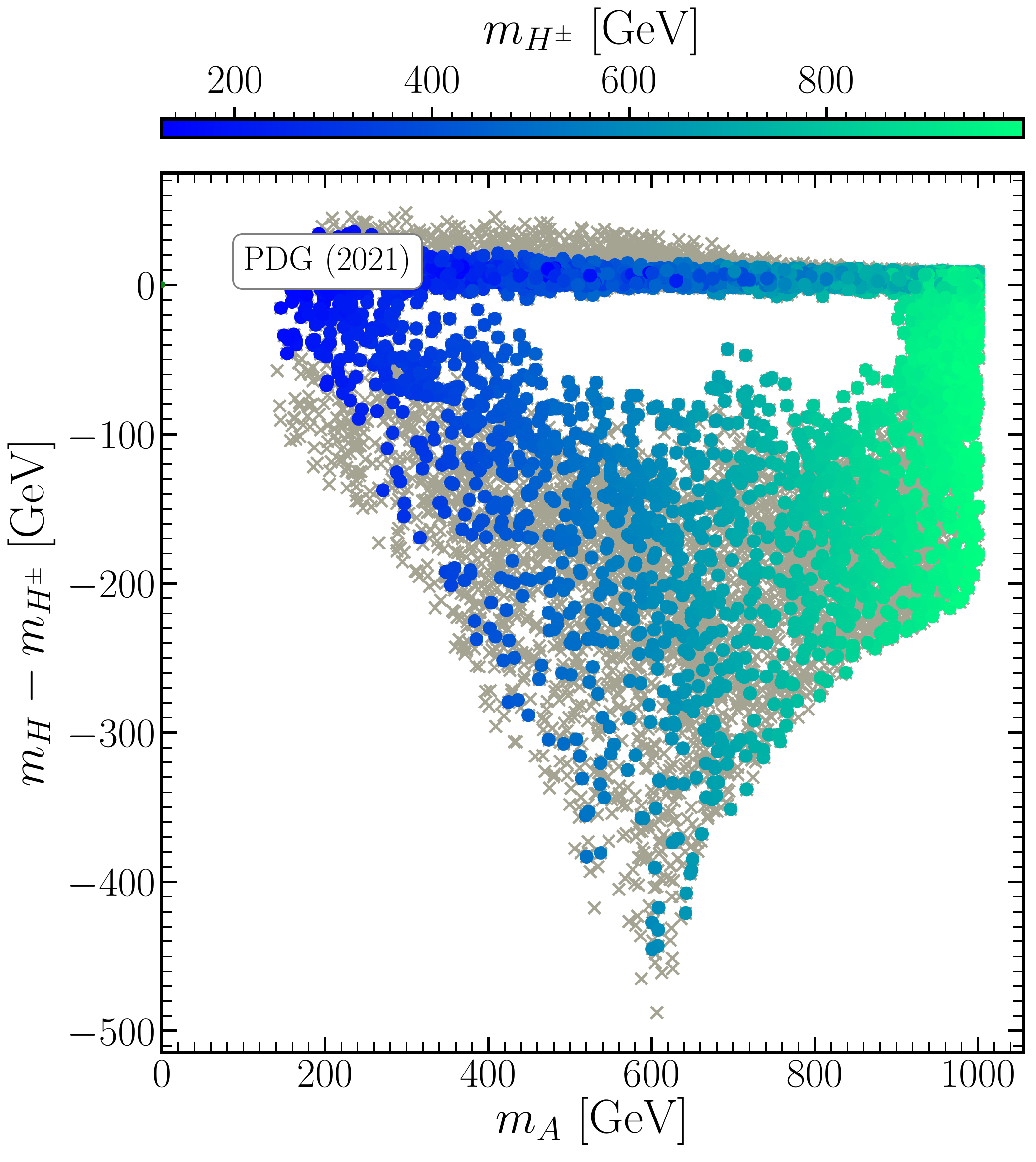}
	\caption{Same as Fig. \ref{fig5} but showcasing PDG-2021 data.}\label{fig6}
\end{figure}
Figs.~\ref{fig3} and \ref{fig4} present the scan results for the EWPOs $S$ and $T$ in the $(m_W, \sin^2\theta_{\mathrm{eff}})$ plane, with the colour gauge indicating $S$  and $T$ values respectively. The light red (gray) band represents the CDF measured value of $m_W$ with its 1$\sigma$ uncertainty, and the orange band represents the world average value for the effective weak mixing angle within its 1$\sigma$ range \cite{ParticleDataGroup:2020ssz,ALEPH:2005ab}. The light blue band signifies the SLD collaboration's most precise measurement of the effective weak mixing angle \cite{ALEPH:2005ab}, also within its 1$\sigma$ uncertainty.

The plot indicates that $m_W$ values near the CDF  measurement align well with the expected range in the 2HDM Type-III, satisfying other theoretical and experimental constraints, including the muon $(g_\mu-2)$ anomaly at the 2$\sigma$ level. Additionally, the $m_W$ values within the CDF's 1$\sigma$ interval correlate well with the SLD reported values but deviate from the world average $\sin^2\theta_{\mathrm{eff}}$.

Interestingly, in this scenario, our model predictions for the new CDF $m_W$ and the muon $(g_\mu-2)$ anomalies receive both negative and positive corrections from $S$, indicated by the colour gauge values of $S\sim -0.02$ and $S\sim 0.06$, while only positive corrections come from $T$. The 2HDM Type-III (IS) effectively reproduces the new CDF $m_W$ measurement and provides a coherent explanation for the muon $(g_\mu-2)$ anomaly.

Given that the mass splitting among $H$, $A$, and $H^\pm$ can influence both $(g_\mu-2)$ and $m_W$, we showcase the scan results in the $(m_A, m_H - m_{H^\pm})$ plane in Figs.~\ref{fig5} and \ref{fig6}. The grey crosses represent parameter sets that meet all established constraints, including the $(g_\mu-2)$ anomaly at the 2$\sigma$ level, yet do not include the EWPOs. In contrast, the coloured points comply with these constraints while also integrating the EWPOs, as indicated by the CDF (2022) data and PDG (2021). A noticeable observation from these results is that compatibility with both the CDF (2022) findings and the $(g_\mu-2)$ data necessitates a non-zero mass splitting among $H$, $A$, and $H^\pm$. Furthermore, the mass differential between $H$ and $H^\pm$ is observed to vary, ranging from a minimum of -439 GeV to a maximum of 45 GeV.
	\subsection{Part II: Normal scenario}
\begin{figure*}[htp!]	
	\centering
	\includegraphics[height=20cm,width=18cm]{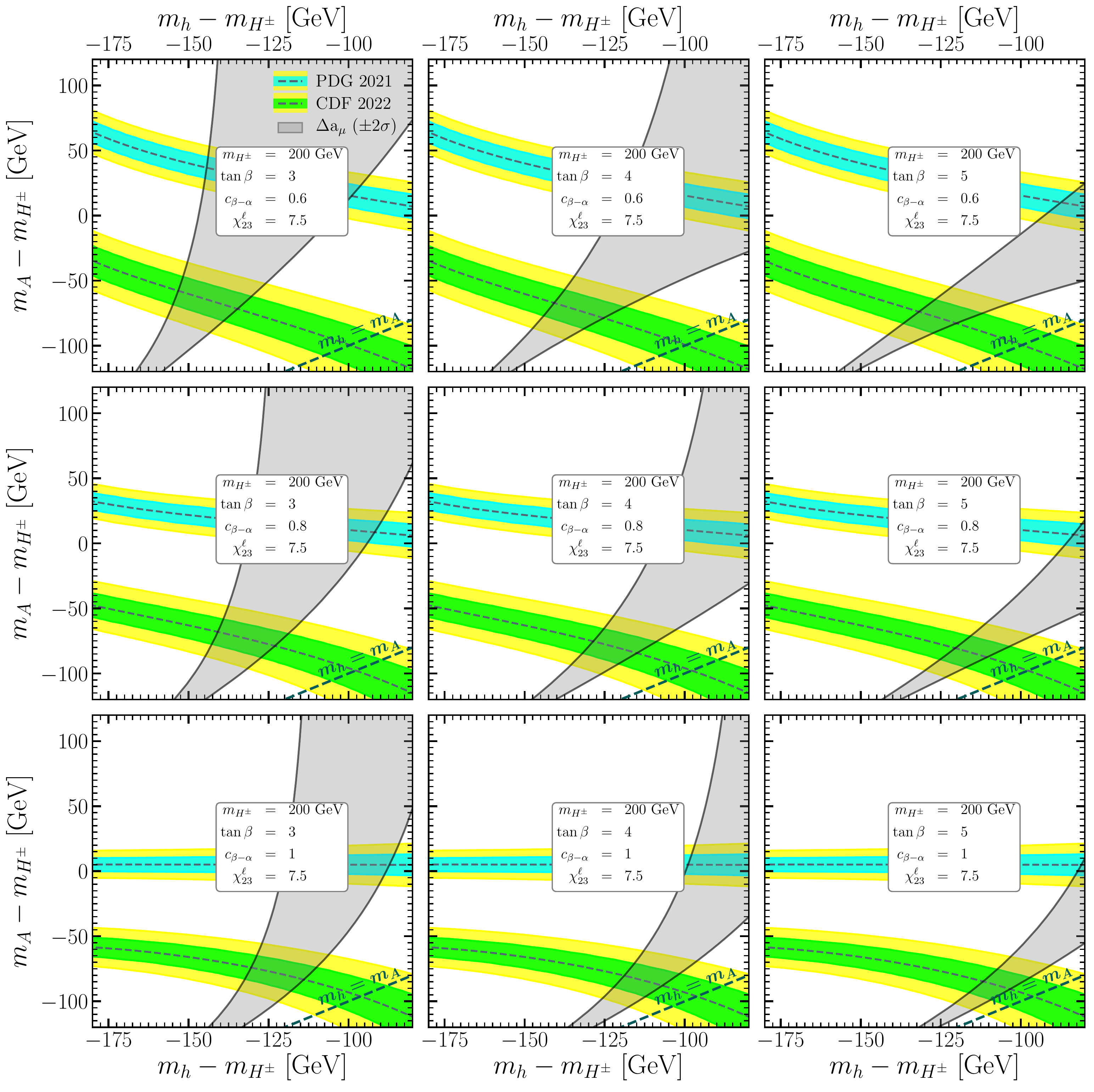}
	\caption{The allowed regions ($1$-and $2\sigma$)  by the oblique parameter fit for 2HDM Type-III in the plane of $m_h-m_{H^\pm}$ versus $m_A-m_{H{^\pm}}$. The gray area shows the $2\sigma$ allowed region by $\Delta a_\mu$. }\label{fig7}
\end{figure*}
\begin{figure}[htp!]
	\centering
	\includegraphics[height=8.5cm,width=8.0cm]{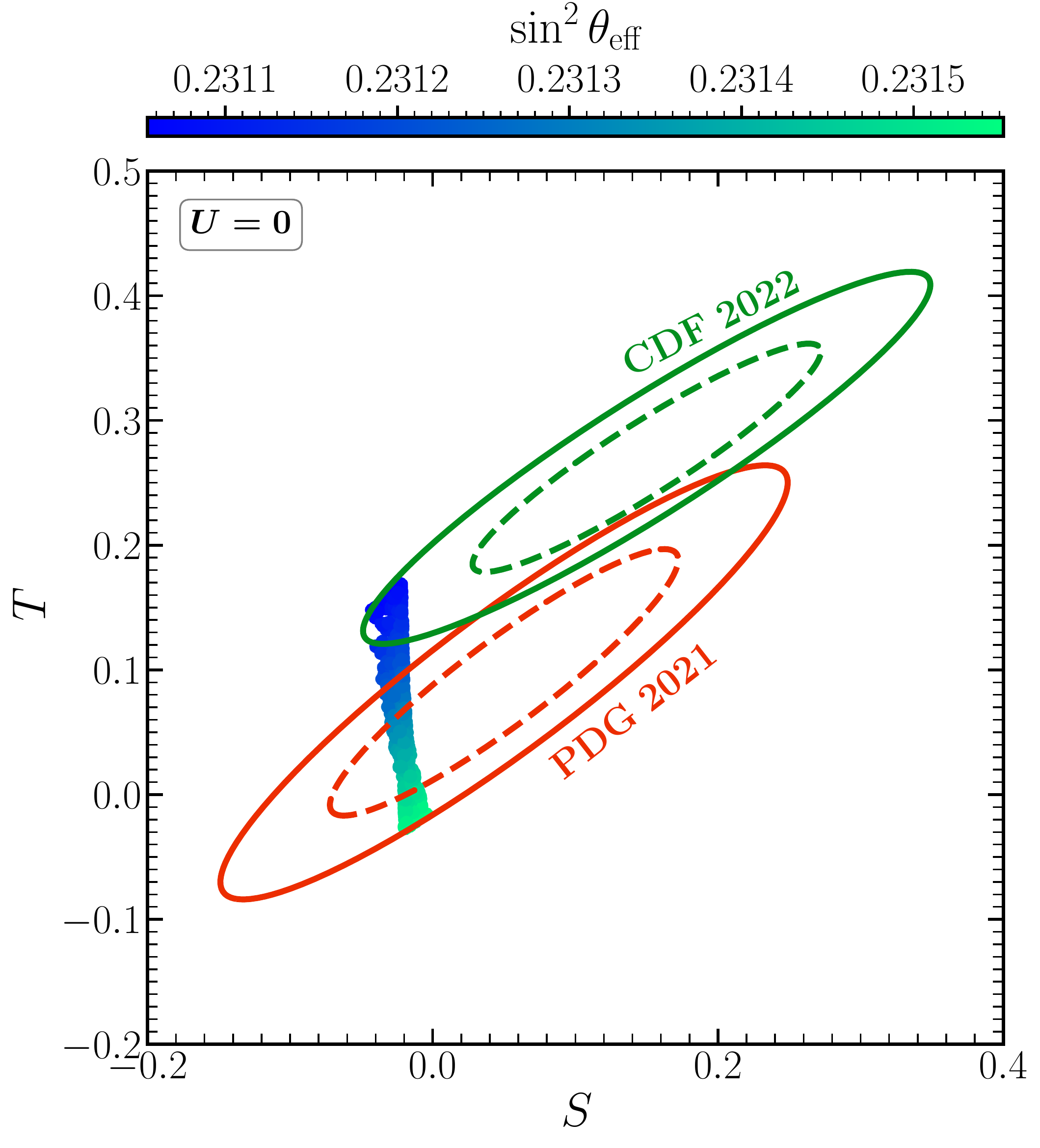}
	\caption{Allowed points by all constraints including $g-2$ superimposed onto the fit limits in the
		$( S, ~ T)$ plane from EWPO data at 68\% (solid lines) and 95\% dashed lines Confidence Level (CL) with a correlation of 92(93)\%  PDG(CDF) with $U=0$. The colour coding indicates the value $\sin^2\theta_{\mathrm{eff}}$.}\label{fig8}
\end{figure}
\begin{figure}[htp!]
	\centering
	\includegraphics[height=8.5cm,width=8.0cm]{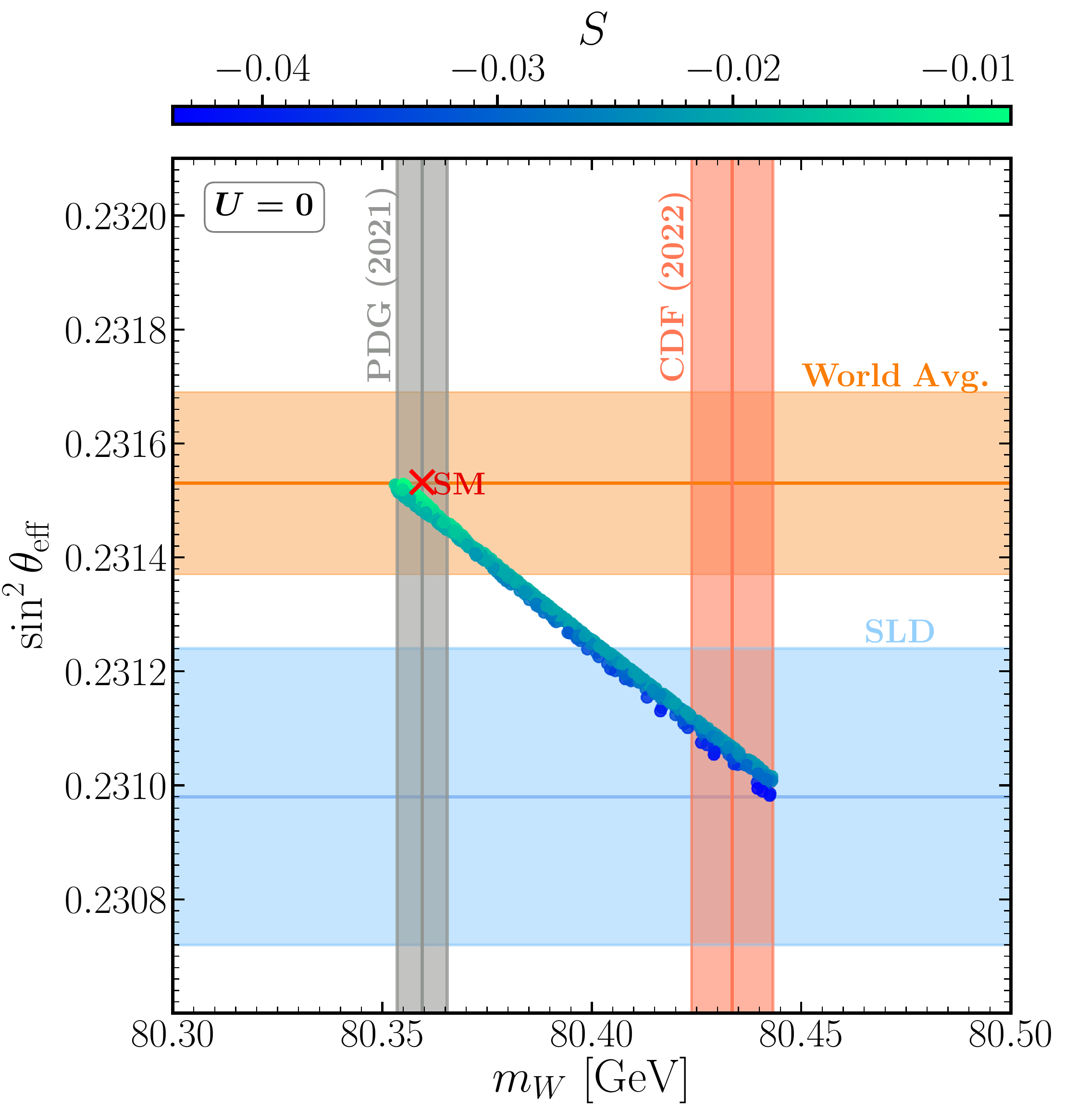}
	\caption{Same points as in Fig. \ref{fig8} mapped into the $(m_W , \sin^2 \theta_{\mathrm{eff}})$ plane. The colour coding indicates the value of $S$. The light-red line represent the CDF-II measured $W$ boson mass and its 1$\sigma$ range (light-red band). The gray line represent the PDG value for the $W$ boson mass with the gray band showing the 1$\sigma$ range. The orange line  represent the world averaged value $0.23153 \pm 0.00016$ \cite{ParticleDataGroup:2020ssz,ALEPH:2005ab} with its 1$\sigma$ range (orange band), while the light blue line illustrates the SLD measured value $0.23098 \pm 0.00026$ \cite{ALEPH:2005ab} of $\sin^2 \theta_{\mathrm{eff}}$ with the light blue band showing the 1$\sigma$ range. The red cross indicates the SM prediction.}\label{fig9}	
\end{figure}
\begin{figure}[htp!]
	\centering
	\includegraphics[height=8.5cm,width=8.0cm]{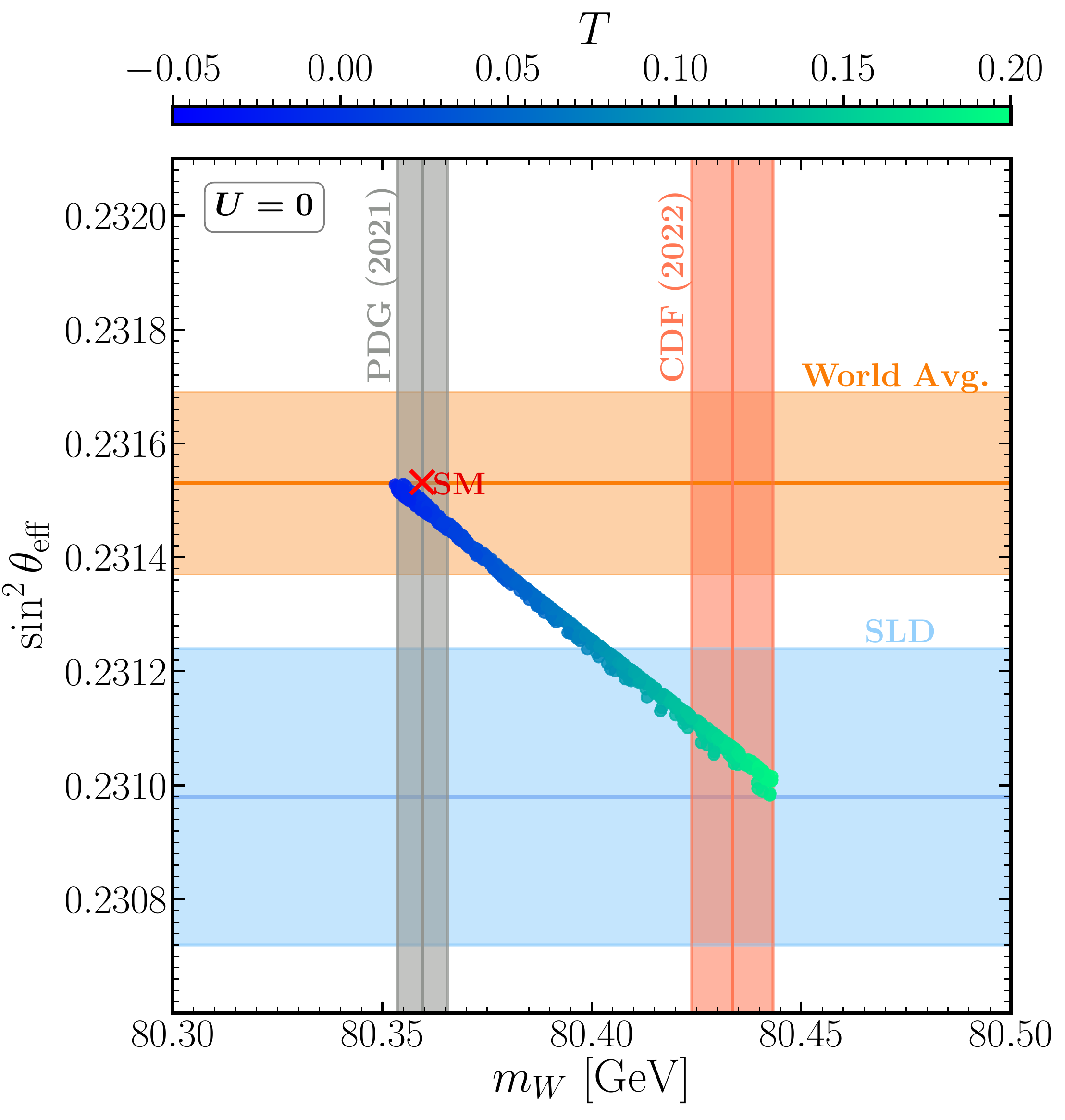}
	\caption{Same as in Fig. \ref{fig6} but with colour bar indicating the $T$ parameter.}\label{fig10}	
\end{figure}
In the normal scenario, we consider the heavy $\mathcal{CP}$-even Higgs boson $H$ to be the Higgs-like particle with a mass of $m_H = 125.09$ GeV. Adhering to a methodology similar to that used previously, we perform a reduced parameter scan with fixed parameters such as $c_{\beta-\alpha}$, $\tan\beta$, and $\chi_{ij}^{f,\ell}$. This scan is represented in Fig.~\ref{fig7}, showing the allowed 1 and 2$\sigma$ regions by PDG (2021) and CDF (2022) in the mass splitting $m_H-m_{H^\pm}$ and $m_H-m_A$ plane, alongside the 2$\sigma$ region permitted by $\Delta a_\mu$. Notably, this phase of the analysis excludes the theoretical and experimental constraints discussed earlier. The figure indicates that mass degeneracy scenarios are disfavoured by the CDF results and the muon anomalous magnetic moment measurements.

Proceeding to the more extensive analysis as outlined in Table~\ref{tab2}, Fig.~\ref{fig8} showcases parameter points that reconcile the muon $(g_\mu-2)$ anomaly within the 2$\sigma$ range, in compliance with further theoretical and experimental constraints. The plot's colour coding, representing $\sin^2\theta_{\mathrm{eff}}$, reveals a distinct range of values. Specifically, for CDF, it lies between [0.23107, 0.23130], and for PDG, between [0.2314, 0.2316]. This indicates that the new CDF $W$-mass measurement can only be explained within the 2$\sigma$ confidence level in this scenario, a finding contrasting with the inverted scenario.

\begin{figure}[htp!]
		\centering
		\includegraphics[height=8.5cm,width=8.0cm]{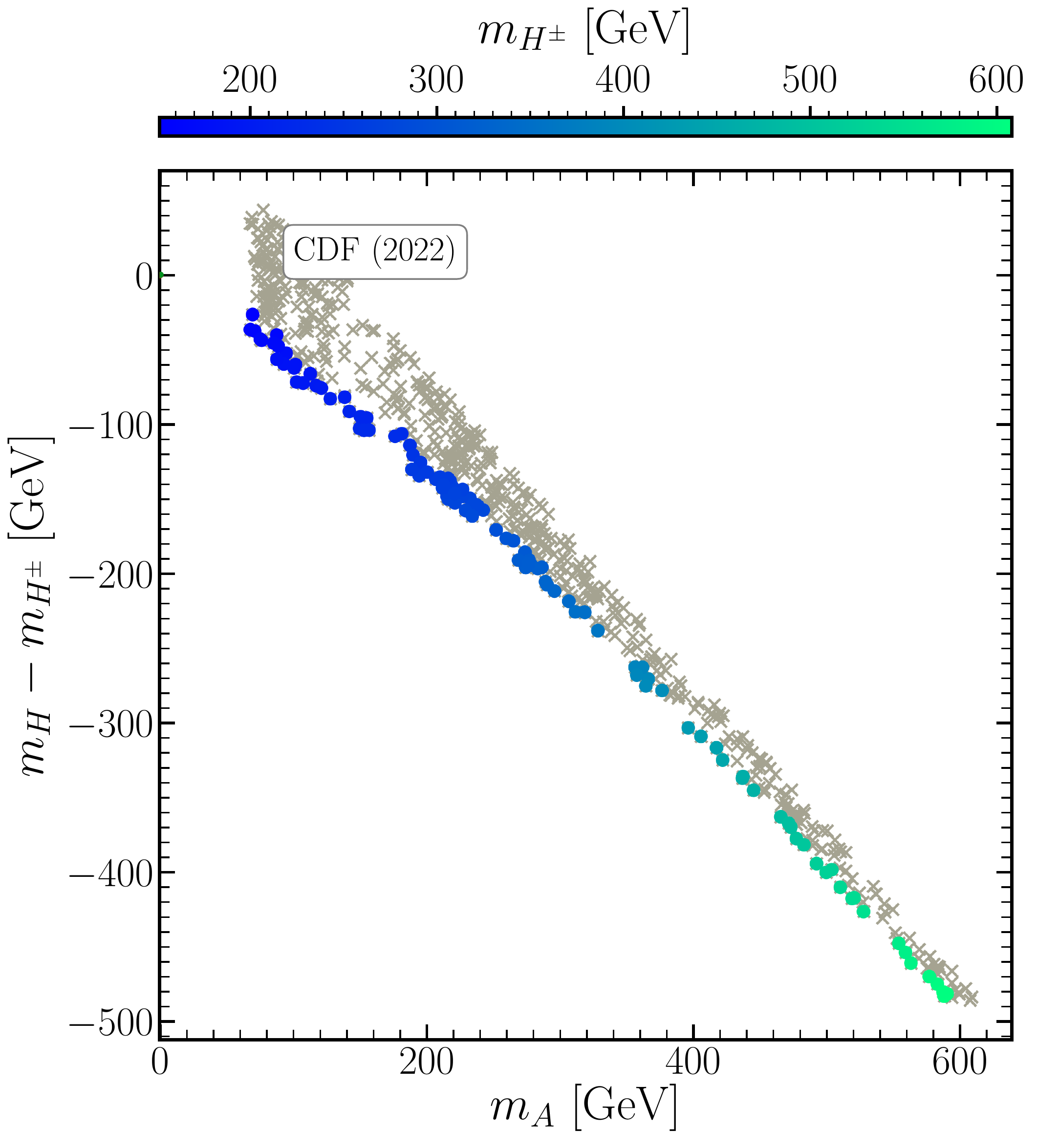}	
	\caption{Same points as in Fig. \ref{fig2} mapped into the $(m_A , m_H-m_{H^\pm})$ plane. The colour coding indicates the value of $m_{H^\pm}$  using the CDF-2022 data. The gray crosses  represent the points allowed only by theoretical and experimental constraints.}\label{fig11}
\end{figure}
\begin{figure}[htp!]
		\centering
		\includegraphics[height=8.5cm,width=8.0cm]{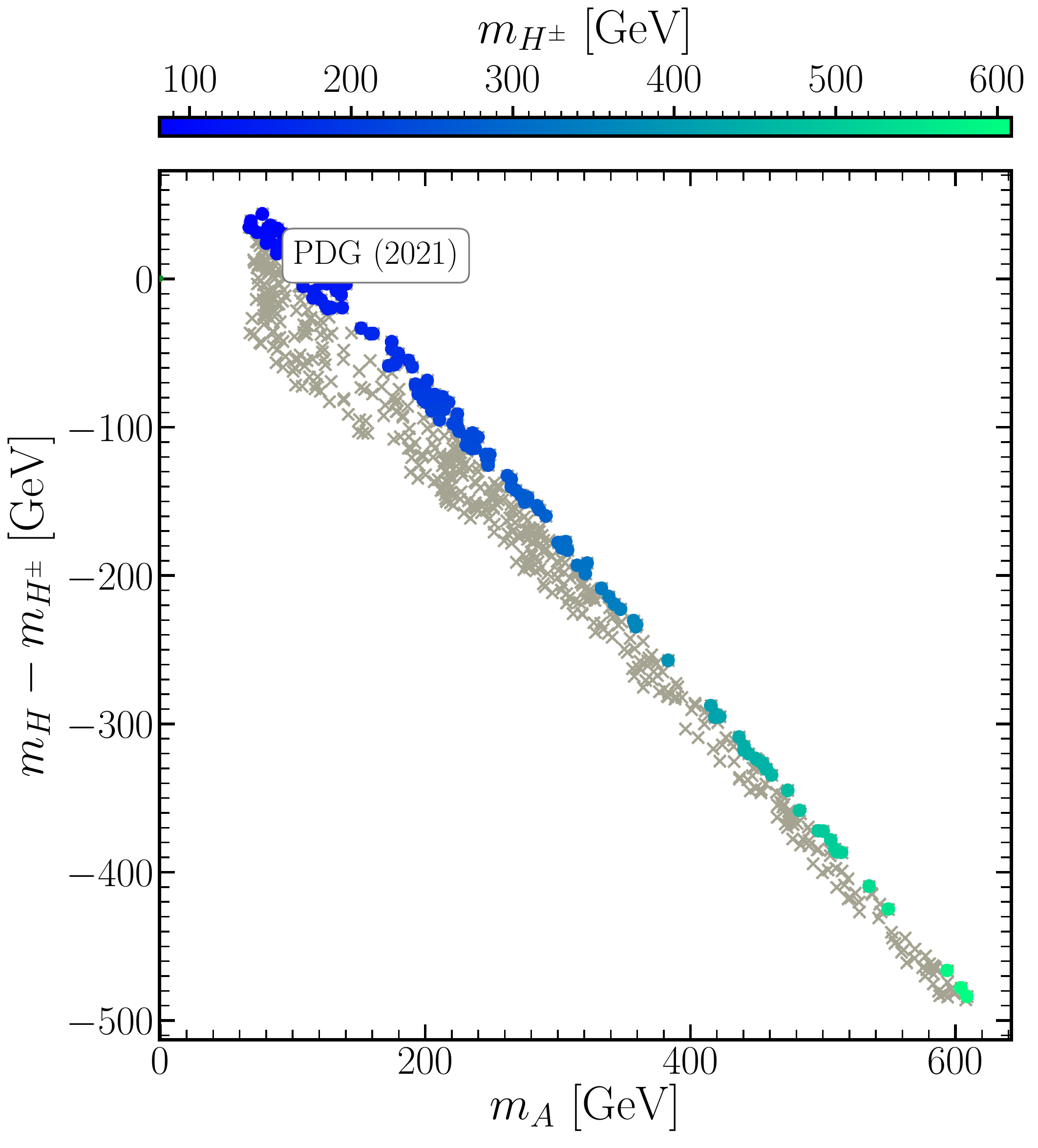}	
	\caption{Same points as Fig. \ref{fig11} but showcasing PDG-2021 data. }\label{fig12}
\end{figure}
Figs.~\ref{fig9} and \ref{fig10} extends this analysis, mapping the same parameter sets onto the $(m_W, \sin^2 \theta_{\mathrm{eff}})$ plane. The vertical red and grey lines, alongside the corresponding bands, represent the CDF and PDG $m_W$ values and their 1$\sigma$ ranges. The horizontal orange and blue lines, with their associated bands, reflect the world average and SLD's measurements of the effective weak mixing angle. This plot underscores a pronounced agreement between the CDF $m_W$ measurement and the SLD value, with only negative $S$ values favoured in this scenario.

Finally, in Figs.~\ref{fig11} and \ref{fig12}, we display our results in the $(m_A, m_h - m_{H^\pm})$ plane. The grey  crosses represent parameter combinations that satisfy all the established constraints, including the $(g_\mu-2)$ anomaly within the 2$\sigma$ level, but do not incorporate EWPOs.In contrast, the colored points are those that meet these constraints and also include the EWPOs, consistent with the data from CDF (2022) and PDG (2021) respectively. Notably, the plots indicates a required negative and non-zero mass difference $m_H - m_{H^\pm}$ to simultaneously accommodate the findings from both $(g_\mu-2)$ and CDF-II.  
\section{Conclusion}

In this work, we investigated the Type III two-Higgs-doublet model (2HDM) to address the recent experimental anomalies in the muon magnetic moment $(g_\mu-2)$ and the $W$-boson mass $m_W$, as reported by FNAL and CDF-II. Our analysis included both normal (NS) and inverted (IS) scenarios in the context of the 125 GeV Higgs boson.

Our study concludes that the 2HDM Type-III, incorporating lepton-flavour violation (LFV), successfully predicts the $m_W$ measurements reported by CDF-II, as well as aligns with the muon $(g_\mu-2)$ data, while remaining consistent with the up-to-date theoretical and experimental constraints. A key observation in both NS and IS scenarios is the preference for non-zero mass splitting among the Higgs bosons $H$, $A$, and $H^\pm$. This outcome, which aligns with the recent $(g_\mu-2)$ and $m_W$ measurements, highlights the impact of mass splittings among these Higgs bosons in reconciling the observed experimental anomalies.

\section{Acknowledgments}

M.Boukidi is grateful for the technical support of CNRST/HPC-MARWAN.

	\bibliography{main}{}
	\bibliographystyle{JHEPcust}

\end{document}